\documentclass[]{spie}

\usepackage{subcaption}
\usepackage{amsmath,amsfonts,amssymb}
\usepackage{graphicx}
\usepackage{hyperref}

\title{Development of TRL5 Space Qualified Hardware for Tuning, Biasing, and Readout of Kilopixel TES Bolometer Arrays}

\author[a]{Graeme Smecher}
\author[b]{Peter Cameron}
\author[c]{Jean-François Cliche}
\author[c]{Matt Dobbs}
\author[c]{Joshua Montgomery}
\affil[a]{Three-Speed Logic, Victoria, Canada}
\affil[b]{Honeywell Aerospace, Ottawa, Canada}
\affil[c]{McGill University, Montreal, Canada}

\authorinfo{E-mail: gsmecher@threespeedlogic.com}

\usepackage{cleveref}
\usepackage{enumitem}
\usepackage{multirow}
\setlist{nosep}

\begin{document}
\maketitle

\begin{abstract}
	The next generation of space-based mm-wave telescopes, such as JAXA's LiteBIRD mission, require focal planes with thousands of detectors in order to achieve their science goals.
	Digital frequency-domain multiplexing (dfmux) techniques allow detector counts to scale without a linear growth in wire harnessing, sub-Kelvin refrigerator loads, and other scaling problems.
	In this paper, we introduce Technology Readiness Level 5 (TRL5) electronics suitable for biasing and readout of LiteBIRD's Transition Edge Sensor (TES) bolometers using dfmux techniques.
	These electronics sit between the spacecraft's payload computer and the cryogenic focal plane, and provide detector biasing, tuning, and readout interfaces between these detectors and the spacecraft's on-board storage.
	We describe the overall architecture of the electronics, including functional decomposition into modules, the numerology and interconnection of these modules, and their internal and external interfaces.
	We describe performance measurements to date, including power consumption, thermal performance, and mass, volume, and reliability estimates.
	This paper is a companion piece to a description of the electronics' on-board Field-Programmable Gate Array (FPGA) firmware.
\end{abstract}


\section{INTRODUCTION}
\label{sec:intro}

Transition Edge Sensors (TESes) are widely used for ground-based observations of the Cosmic Microwave Background.
Their success in experimental cosmology reflects both the exquisite sensitivity of these devices at microwave frequencies, as well as the technological maturity of TES readout systems as a whole.
Over the past decade, the scale of TES arrays deployed for cosmology has increased by over an order of magnitude, with SPT-3G currently fielding an array of 15,000 TESes within a single focal plane.

The rapid pace of development for ground-based TES readout systems heralds adoption of the technology in space.
Several proposed space observatories have baselined large arrays of TES detectors for microwave, infrared, or X-ray missions.
The first of these missions is JAXA's LiteBIRD Space Observatory, scheduled to launch by 2029.

Although ground-based technologies have progressed into sophisticated and reliable systems, there is much work still to be done in qualifying such systems for launch and use in space environments.
One important aspect of this effort is the development of space-qualified control electronics, capable of performing the multiplexing operations required to operate large TES arrays.

Multiplexing systems have been an enabling technology for ground-based observatories.
The most common systems fall into two categories: time-division multiplexing (TDM), and frequency domain multiplexing (FDM).
The system described here implements digital frequency-domain multiplexing (dfmux)\cite{Smecher2022b}.
Each TES in an array is operated with distinct AC bias voltages between 1-10 MHz.
Incident power is recorded through amplitude modulation of these carrier sinusoids, in an operation analogous to AM radio.
A description of the associated firmware and signal processing pipeline is given in a companion paper (Ref.~\citenum{Smecher2022b}).

In this document, we describe the overall design of a TRL-5 readout system implementing power-efficient, highly-multiplexed FDM.
This system is designed to be composable to scale across a variety of space-based TES applications.
It has been selected as a baseline for the LiteBIRD Space Observatory.

\section{Concept of Operations}
\label{sec:conops}

The readout system is the interface between the cryogenic circuitry and the digital electronics that transfer and store data for offline analysis.
The readout system has two functions:
\begin{itemize}
	\item
		to ``tune'' the cryogenic hardware (detectors and cryogenic amplifiers) into the appropriate state to take data; and
	\item
		to digitize and process focal-plane signals into a form suitable for transfer, storage, and offline analysis.
\end{itemize}

A third essential requirement, which becomes acute in space, is uninterrupted operation in spite of environmental conditions including radiation.
Because TES detectors operate in their superconducting transition, they are prevented from entering the super-conducting state through precisely calibrated application of electrical power from the readout.
Interruptions to that bias power can dramatically alter the state of the cryogenic hardware.
Recovery from interruptions require a series of tuning operations before the system is fit to resume taking data.
Rapid and efficient fault detection and recovery is an active area of investigation at all levels of mission design, including detector fabrication, readout electronics design, and system integration.

\subsection{Architectural Overview}

\begin{figure}
	\centering
	\subcaptionbox{
		Annotated cutaway view of an SPU, showing internal assemblies.
		Each SPU contains its own Power Conditioning Assembly (PCA), Signal Processing Assembly (SPA), and up to 4 Digitizer Assemblies (DAs), which are all mechanically and thermally contained within a Signal Processing Unit Enclosure (SPUE).
		\label{fig:spu-cutaway-annotated}
	}{\includegraphics[width=0.5\textwidth]{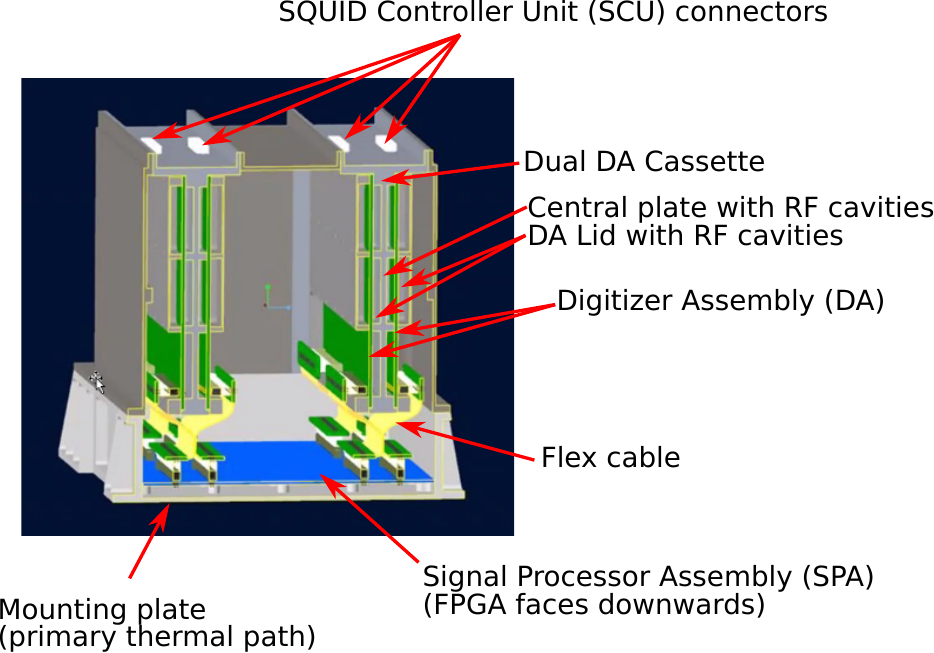}}%
	\hfill
	\subcaptionbox{
		An SPU.
		A cover of the enclosure has been removed, showing one of four Digitizer Assemblies (DAs).
		The Signal Processing Assembly (SPA) occupies the bottom cavity, allowing efficient thermal coupling of the FPGA to the attachment surface (bottom).
		\label{fig:spue}
	}[0.45\textwidth]{\includegraphics[width=0.3\textwidth]{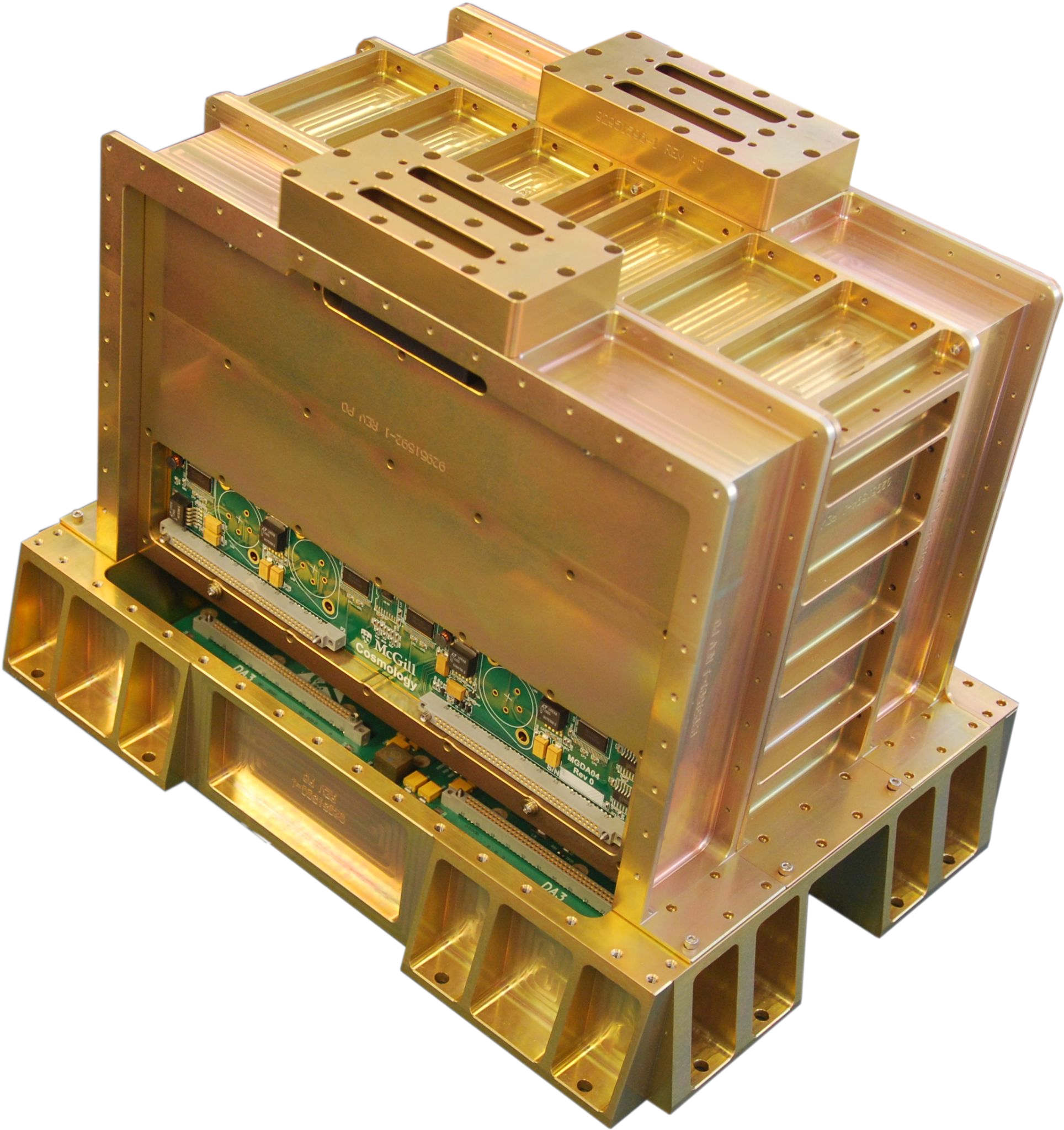}}
	\caption{A Signal Processing Unit (SPU).}
\end{figure}

The overall architecture of the readout system matches the ground-based ``ICE'' system\cite{bandura2016ice} and previous development work on a space-qualifiable system\cite{Bender2014}.
A single modular unit is known as a Signal Processing Unit (SPU).
A cutaway view of a single SPU is provided in \Cref{fig:spu-cutaway-annotated}.
A partially integrated SPU is shown in \Cref{fig:spue}.
Each SPU contains its own Power Conditioning Assembly (PCA), Signal Processing Assembly (SPA), and up to 4 Digitizer Assemblies (DAs).
These assemblies are mechanically and thermally contained within a Signal Processing Unit Enclosure (SPUE).
Each SPU can operate 16 SQUID channels at a multiplexing factor of up to 128 detectors per SQUID.
Signal processing firmware operating on an FPGA within each SPU is described in a companion paper\cite{Smecher2022b}.

Between the SPU and focal plane are SQUID Controller Assemblies (SCAs) enclosed within SQUID Controller Units (SCUs).
SCAs interface directly to the cryogenic stage, performing amplification of analog signals and conditioning of the bias voltages and nulling currents.
Both SCAs and DAs accommodate up to 4 multiplexed SQUID channels.
For the LiteBIRD mission, a flight-qualified SCA based on an earlier TRL-5 design\cite{Bender2014} is being built by the National Institute for Nuclear Physics (INFN) in Italy.
Although all tests reported in this document use a ground-based SCA (MGNGSQ4), the two SQUID controller designs are functionally equivalent.

\subsection{System Scalability}

Each SPU is an independent, modular unit.
It can be configured as either a primary or redundant system, and may be populated with up to 4 DAs.
This flexibility allows the system to scale in size without lowering the overall TRL.

\begin{figure}
	\centering
	\includegraphics[width=\textwidth]{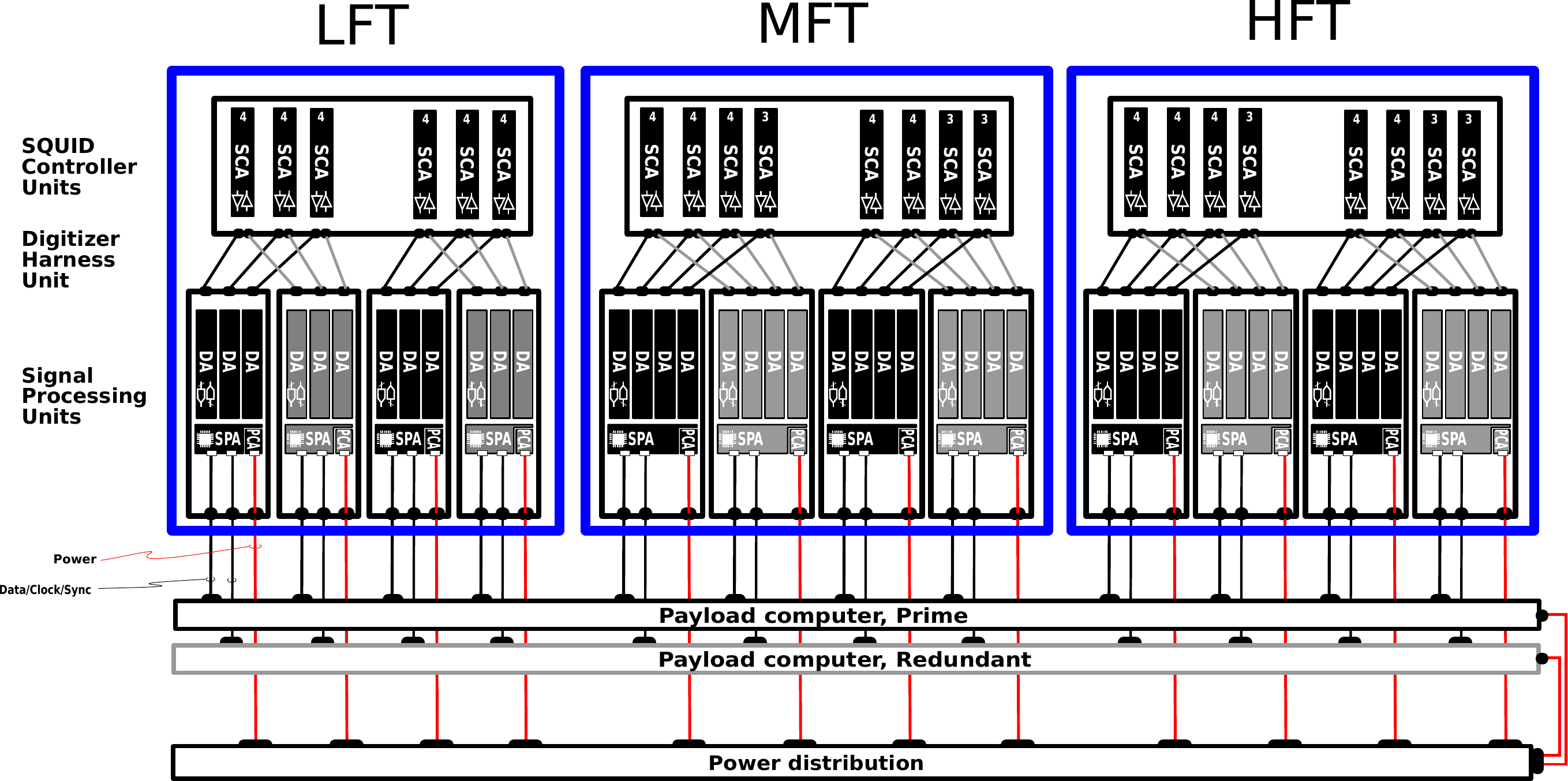}
	\vspace{2ex}
	\caption{
		The architecture and numerology of LiteBIRD's readout electronics.
		There are 6 active and 6 cold-spare SPUs communicating with 82 SQUIDs split across 3 telescopes (LFT, MFT, and HFT).
		Cross-strapping between primary and redundant units occurs between SCUs and SPUs, and between SPUs and payload computers (DPUs).
		In LiteBIRD, redundancy is provided by cold-spare SPUs and DPUs.
	}
	\label{fig:system-block-diagram}
\end{figure}

\Cref{fig:system-block-diagram} shows the readout system as configured for the LiteBIRD mission.
Three separate telescopes (Low-Frequency Telescope, Mid-Frequency Telescope, and High-Frequency Telescope) house a combined 82 SQUIDs and approximately 4,500 TES bolometers.
To control them, a total of 12 SPUs are required: 6 primary, and 6 cold spares.
Each SCA has two DA interfaces: one primary, and one cold spare.
In this configuration, the SCAs are the only element of the warm readout electronics with no cold-spares -- they are also the lowest-power element, have limited digital circuitry, and a low integrated risk of failure (see \Cref{sec:reliability} for a discussion on reliability).

Various SQUID-focal-plane mappings can be accommodated by this design.
Although the warm readout electronics can multiplex as high as 128 detectors per SQUID, the LiteBIRD configuration doesn't use a multiplexing factor higher than 68.
As the LiteBIRD focal-plane has matured, the optimal readout electronics numerology has evolved.
We have accommodated these changes by optimizing the multiplexing factor assigned for each telescope and the number of DAs populated in each SPU.
These changes have no TRL impact, and require no additional R\&D or qualification.

\subsection{Detector Tuning and Operation}

The high-level algorithms that tune a detector array involve complex interactions between cryogenic elements (bolometers, SQUIDs) and electronics that control them (SPUs, SCUs, and active elements in the cooling chain).
These algorithms are implemented in a Data Processing Unit (DPU), which performs top-level telescope control operations and handles data streamed from SPUs to on-board storage.
The DPU is outside the scope of hardware described here.

Primary and redundant DPUs communicate with SPUs through point-to-point, bidirectional serial links.
Each link is full-duplex and has an usable bandwidth of 40 Mbit/s accounting for encoding overhead.
(See Ref.~\citenum{Smecher2022b}.)

After power-up, SPUs must be configured by a DPU to initialize DAs and SCAs, and to assign bias parameters (e.g. frequencies, phases, amplitudes) required by the dfmux signal paths.
All DPU-to-SPU commands consist of reads and writes into memory-mapped registers in the SPA's FPGA.

Once an SPU is active, it continuously streams data for all 128 multiplexed channels on all 16 SQUIDs.
This data accounts for about 20 Mbits/s of the available bandwidth from the SPU to the DPU.
The remaining bandwidth is available for commanding and data returns associated with tuning operations, and is wide enough not to be a limiting factor in tuning speed.

\section{SPU Assemblies}

Below we briefly describe the designs of the major elements of the SPU: the SPA, DA, and SPUE.

\subsection{Signal Processing Assembly (SPA)}

\begin{figure}
	\centering
	\includegraphics[width=\textwidth]{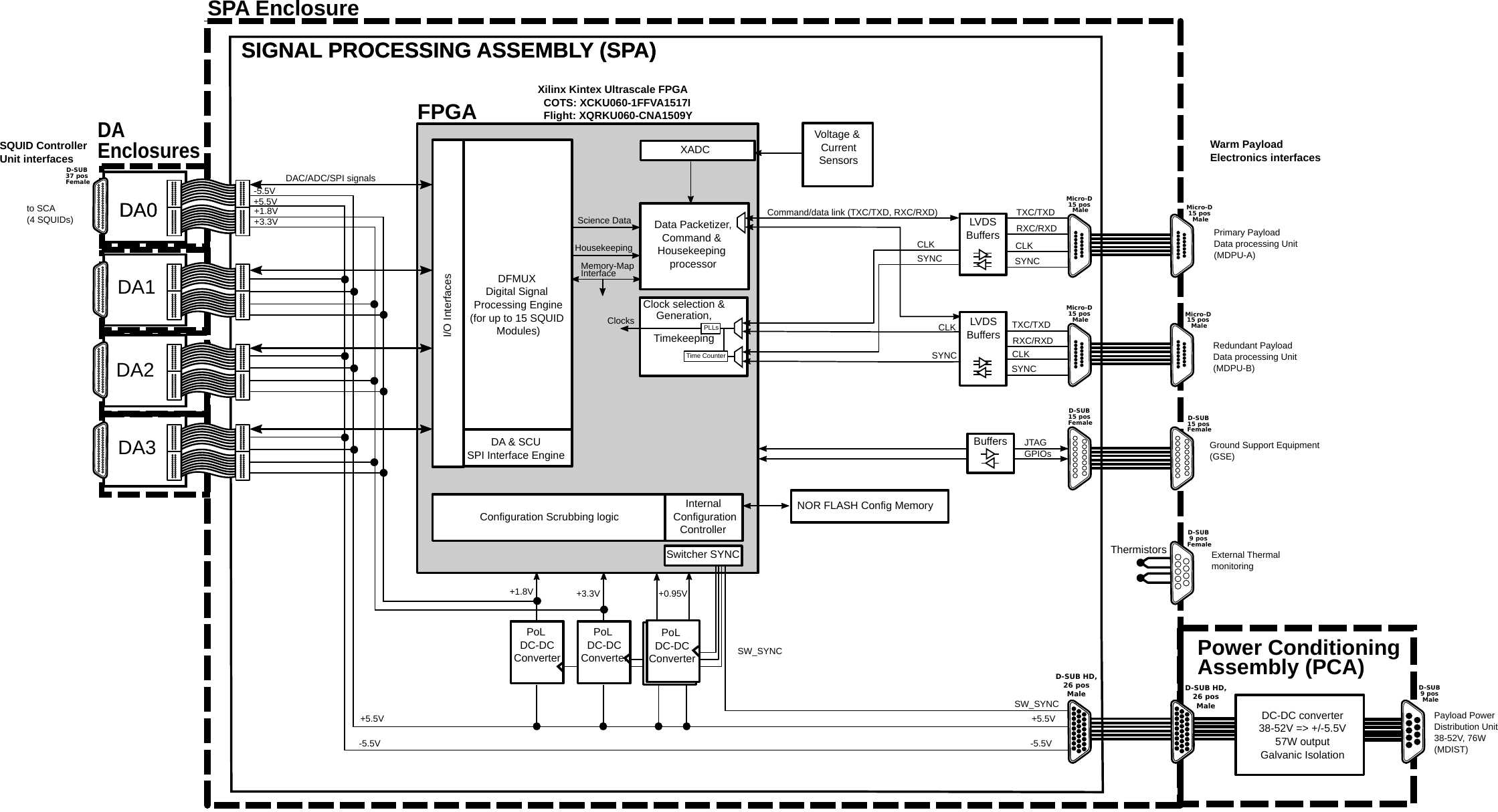}
	\vspace{2ex}
	\caption{
		Signal Processing Assembly (SPA) block diagram.
	}
	\label{fig:spa-block-diagram}
\end{figure}

The SPA executes the real-time synthesis and demodulation algorithms that are central to a dfmux readout system.
It consists of an FPGA (XQRKU060) and associated power supply, interface, and support circuitry allowing it to operate.
The FPGA is mounted on the PCB's underside and bonded directly to the enclosure to facilitate a short thermal path to radiators on the spacecraft.
A block diagram of the SPA is shown in \Cref{fig:spa-block-diagram}.

\begin{figure}
	\centering
	\hfill
	\subcaptionbox{
		A Signal Processing Assembly (SPA).
		The FPGA is mounted on the underside of the board, facilitating a good connection to the thermal interface on the chassis (see \Cref{fig:spue}).
		The 8 large, white connectors attach to 4 DAs via flexible adapters.
		Other ports (2 DPU connectors, 1 Ground-Support Equipment (GSE) connector, and 1 PCA connector) attach to bulkhead interfaces on the chassis via short ``pigtail'' cables.
		A block diagram of the SPA is shown in \Cref{fig:spa-block-diagram}.
		\label{fig:spa}
	}{\includegraphics[width=0.45\textwidth]{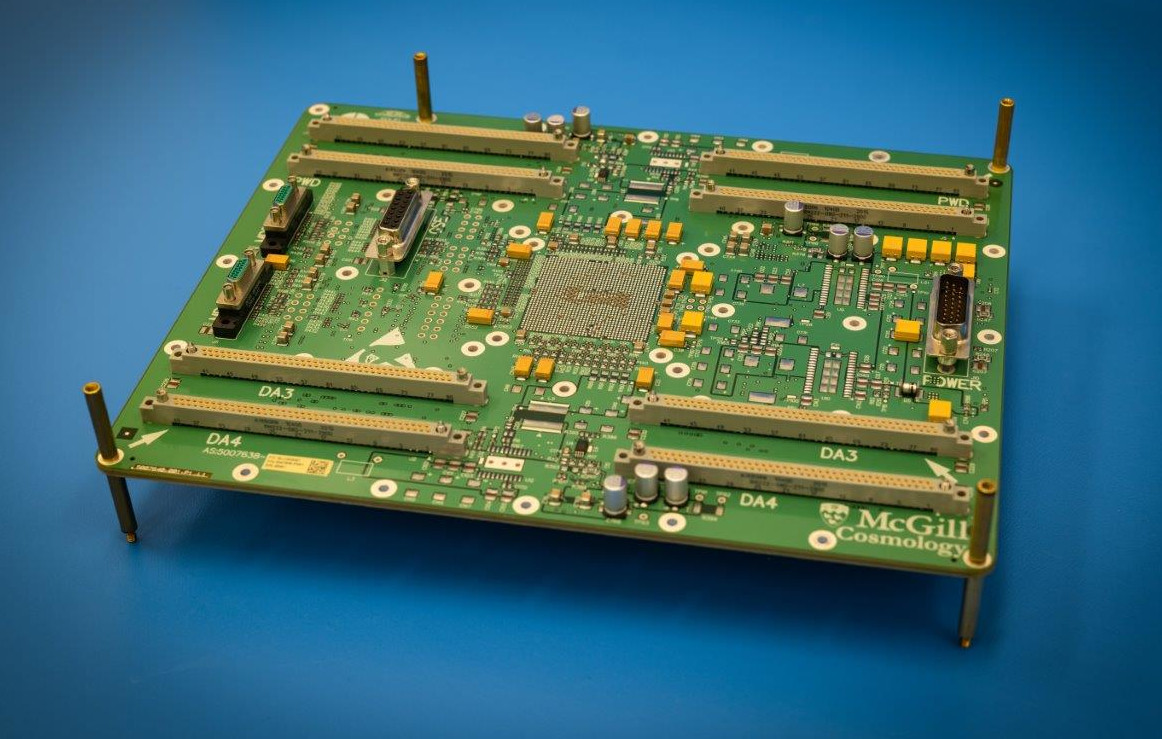}}%
	\hfill
	\subcaptionbox{
		A Digitizer Assembly (DA).
		Each DA contains carrier, nuller, and demodulator ADCs/DACs for 4 SQUID channels, plus associated amplifiers, filters, and housekeeping.
		A flexible harness (top) connects to the SPA.
		A connector (bottom) interfaces to the connector at the bottom via an external wire harness.
		A block diagram of the DA is shown in \Cref{fig:da-block-diagram}.
		\label{fig:da}
	}{\includegraphics[width=0.45\textwidth]{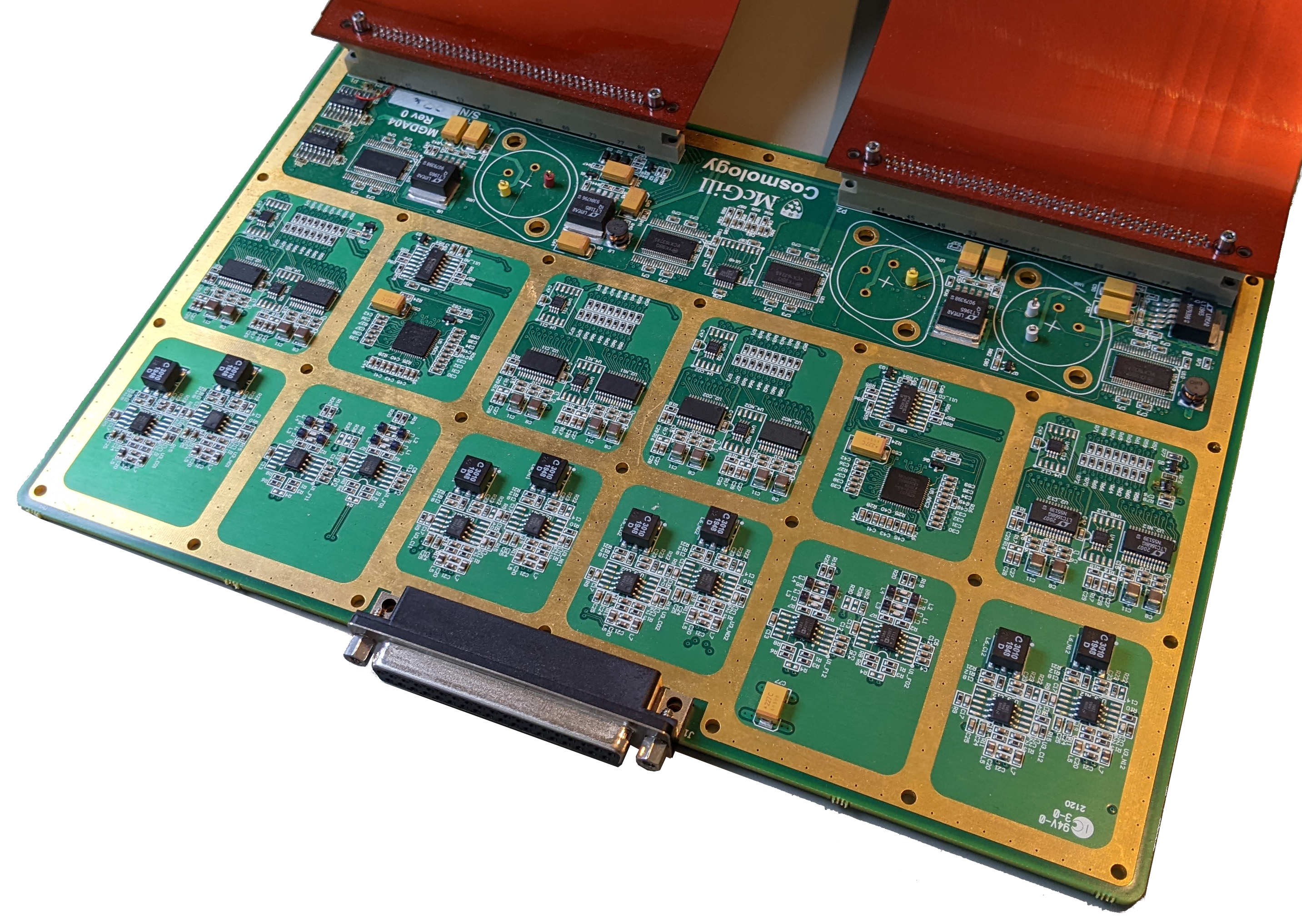}}%
	\hfill
	\vspace{2ex}
	\caption{
		Major assemblies within the Signal Processing Unit (SPU).
		These assemblies' orientations within an assembled SPU can be seen in \Cref{fig:spu-cutaway-annotated}.}
\end{figure}

All oscillators in the system are synchronized to a single, externally-supplied clock to ensure EMI throughout the system is controlled.
The sole exception is an oscillator within the FPGA\cite{Xilinx-ug570}, which is unavoidable but not used or exposed outside the FPGA.
Point-of-load regulators in the SPA operate at 500 kHz, and are phase-locked to the 10 MHz input clock.
When the SPA is powered up, all converters free-run around their nominal frequencies until the FPGA firmware is fully loaded, at which time it provides the synchronization signal for each converter.

To minimize power consumption and design complexity, no external RAM is present on the SPA.
The only storage elements are the non-volatile flash containing the FPGA's bitstream, and volatile memory inside the FPGA itself.

\subsection{Digitizer Assembly (DA)}
\label{sec:da}

\begin{figure}
	\centering
	\includegraphics[width=0.6\textwidth]{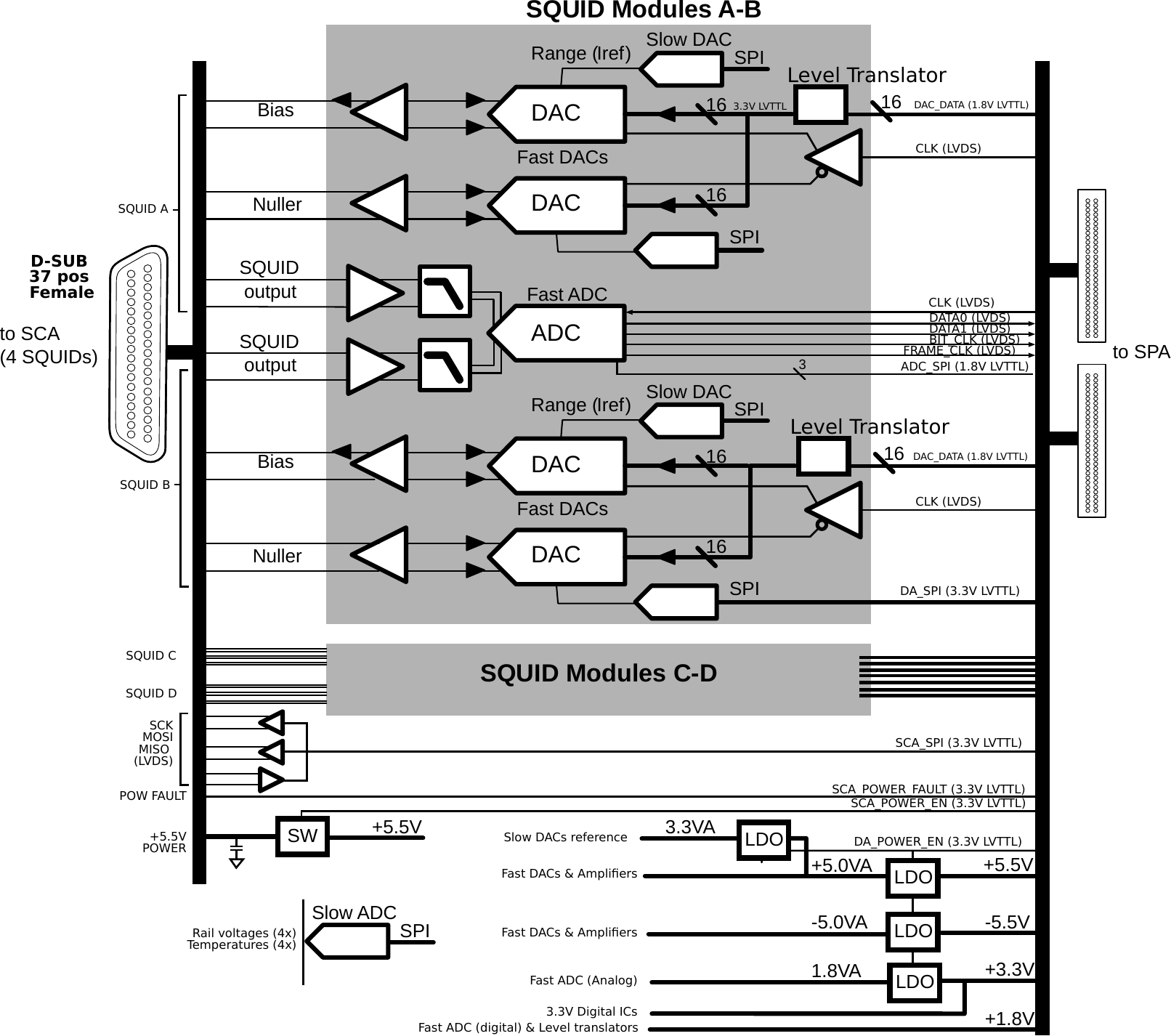}
	\vspace{2ex}
	\caption{
		Digitizer Assembly (DA) block diagram.
		Each Signal Processing Unit (SPU) accommodates up to 4 DAs.
		Each DA contains carrier DACs, nuller DACs, and demodulator ADCs sufficient for 4 SQUID channels.
		These fast DACs operate at 20 MSPS.
		Each DA attaches via a cable harness to a SQUID Controller Assembly (SCA), and via a flexible harness to a Signal Processing Assembly (SPA).
	}
	\label{fig:da-block-diagram}
\end{figure}

The Digitizer Assembly (DA) houses ADCs, DACs, filters, and amplifiers that interface between the analog signals on the focal plane and the digital algorithms executed in the SPA.
Each SPA connects to up to 4 DAs via flexible harnesses.
Each DA attaches via a 37-pin D-subminiature wire harness to an SCA.
A block diagram of the DA is shown in \Cref{fig:da-block-diagram}.
A prototype DA is shown in \Cref{fig:da}.

Neither the fast ADC nor DAC come from vendors' radiation-qualified product lines.
(Radiation-hard parts exist but are not competitive in terms of noise performance, power consumption, or channel density.)
Both the ADC and DAC have been radiation tested.
Test results are favourable, though a detailed analysis is ongoing.

\subsection{Signal Processing Unit Enclosure (SPUE)}

The SPU Enclosure is shown in \Cref{fig:spue}.
The enclosure:
\begin{itemize}
	\item protects the delicate column grid array (CGA) of the FPGA,
	\item maintains rigidity for launch stresses,
	\item conducts heat away from the FPGA and other highly dissipative components, and
	\item fully encloses the system to prevent electromagnetic interference.
\end{itemize}

The SPUE is constructed from gold-plated aluminum.

\subsection{Power Converter Assembly (PCA)}

The Power Converter Assembly (PCA) provides an isolated power-conversion stage between unregulated bus supplies and the internal supplies used by the SPA and DA.
The PCA also supplies power to SCAs attached to each SPU.
The PCA is synchronized to a clock supplied by the SPA.

PCAs are off-the-shelf units, with varying power-conversion efficiencies (typically, 70\%-90\% depending on power-conversion ratio).
For LiteBIRD, the spacecraft supply voltage is between 38-52 V and we assume a nominal PCA efficiency of 72\%.

\section{Performance}
\label{sec:results}

Detailed analysis and testing, including thermal-vacuum (TVAC), launch-load, and shake-table testing was completed during a recent technology development (STDP) contract from the Canadian Space Agency (CSA).
The SPU meets LiteBIRD's current requirements.
A summary of analyses and/or tests to date is provided in \Cref{tab:test-results}.

\begin{table}
	\centering
	\begin{tabular}{l|l}
		\hline
		\textbf{Parameter} & \textbf{Value} \\ \hline
		Mass & 9.5 kg \\
		Volume & 34cm x 33cm x 27cm \\
		Lowest Eigenfrequency (simulation) & 466 Hz \\
		Quasistatic Launch Load (simulation) & 55G \\
		Random vibration (in-plane) & 9.9 Grms \\
		Random vibration (out-of-plane) & 15.2 Grms \\
		Power Consumption & 43 W \\
		Multiplexing Factor & up to 128× \\
		Total Supported TES channels & 2,048 \\
		Survival Temperature & -50°C - 100°C \\
		Operational Temperature & -50°C - 76°C \\ \hline
	\end{tabular}
	\vspace{2ex}
	\caption{
		Key Parameters for a single SPU, excluding the PCA (which is expected to be a COTS unit).
		Results not marked ``simulation'' represent tests/measurements.
		Power consumption is extrapolated from 1 DA and 1 SCA to 4 of each, and includes losses assuming 72\% PCA efficiency.
		The baseline LiteBIRD readout uses 6 active SPUs.
	}
	\label{tab:test-results}
\end{table}

Below, we provide details associated with power consumption, thermal dissipation, mass, and reliability.
Performance and validation activities were performed using a COTS power supply for the PCA and the ground-based SCA that is part of the ICE readout system.\cite{Bender2014, bandura2016ice}.
Noise and tuning benchmarks with the INFN SCA and flight-representative cryogenic hardware will follow.
For firmware performance, see Ref.\citenum{Smecher2022b}.

\subsection{Power consumption}

A fully populated SPU dissipates 31 W under nominal conditions at room temperature (25°C) and excluding PCA losses.
This SPU must be coupled with 4 SCAs (1.8W each for the INFN design; 2.5W each for the ground-based design, excluding PCA losses) to read out 16 SQUID channels.
With 2.5W per SCA and a PCA efficiency of 72\% PCA efficiency, a complete readout system for 16 SQUIDs at a multiplexing factor of up to 128× dissipates 57 W (84 mW/bolometer at 68×; 45mW/bolometer at 128× multiplexing).
The SPA's power consumption was measured at a full 128× multiplexing factor and decreases at lower multiplexing factors.

The LiteBIRD configuration operates 6 SPU/SCU pairs (the other 6 SPUs are cold spare).
The 2 LFT SPU/SCU pairs have only 3 DAs and 3 SCAs each.
The 2 MFT and 2 HFT SPU/SCU pairs are fully populated.
The total configuration for LiteBIRD dissipates 324 W.
With LiteBIRD's baseline flight configuration (4,500 TESes), power dissipation is 72 mW/TES.
(This hardware configuration is capable of reading out 11,264 TESes at 128× multiplexing, corresponding to a per-TES power consumption of 29 mW/TES.)
A detailed breakdown of power consumption is given in \Cref{tab:power-consumption}.

\begin{table}
	\centering
	\begin{tabular}[H]{l|l}
		\hline
		\textbf{Parameter} & \textbf{Value} \\ \hline
		1 SPA, 128× multiplexing & 15.4 W\\
		1 DA & 3.9 W \\
		%1 INFN SCA & 1.8 W \\
		1 ground-based SCA & 2.5 W \\
		PCA Efficiency & 72\% \\ \hline
		LFT SPU, (3 DAs, 3 SCAs) & 48 W \\
		MFT/HFT SPU, (4 DAs, 4 SCAs) & 57 W \\ \hline
		\textbf{LiteBIRD configuration, total (6 SPUs)} & \textbf{324 W} \\
		%Total LiteBIRD TESes & 4,500 \\ \hline
		\textbf{LiteBIRD configuration (4500 TESes)} & \textbf{72 mW/TES} \\ \hline
	\end{tabular}
	\vspace{2ex}
	\caption{
		Power consumption for architectural elements and scenarios.
		These are room-temperature measurements, not specifications, and do not include design margin.
		Unlike Ref.~\citenum{Bender2014}, these figures are inclusive of power-converter losses.
		Although the INFN SCA is more power-efficient than the ground-based SCA, we use the more pessimistic figures here to match testing completed to date.
	}
	\label{tab:power-consumption}
\end{table}

\subsection{Thermal dissipation}

About 80\% (43 W) of the power supplied to the SPU is dissipated as heat within the SPU itself.
(The remainder is dissipated in the SCAs powered by the SPU.)

The SPU manages thermal dissipation from the readout with only conductive cooling.
The FPGA is the single largest source of thermal dissipation, and is mounted on the underside of the SPA.
The FPGA is bonded directly to the SPUE chassis, providing both mechanical and thermal support.
The total thermal path between FPGA die and spacecraft radiators is only a few centimeters and has a temperature gradient of approximately 10°C.
For the flight FPGA and flight bonding materials, the simulated thermal gradient is 7°C.
Measurements in a TVAC chamber using demonstration hardware showed a 14°C gradient, consistent with the material and configuration differences.
The maximum operating temperature (before derating) of the flight FPGA is 125°C, which provides substantial margin for radiator capacity.
(Current requirements on the LiteBIRD radiator deck specify an operational maximum of 40°C.)

The DA cassettes are also designed to conduct heat towards the radiator deck, but the dissipative sources are more distributed.
Most of the SPUE structure is in service of rigidity rather than thermal management.

\subsection{Mass}

An assembled SPU has a mass of 9.5 kg, excluding the PCA.
Both the structural and thermal performance of the SPUE exceed target requirements.
A preliminary analysis suggests it is possible to reduce the mass by 20\% without degrading structural or thermal performance.

\subsection{Reliability}
\label{sec:reliability}

In a CMB mission such as LiteBIRD, science products require long-term data collection over wide angular scales.
A given map of the CMB is formed numerically by combining many sweeps of each telescope's beam, which is constantly in motion, across the region of interest.
This approach is in sharp contrast to telescopes operating at small angular scales and producing short-term images of specific targets.
It is consistent with ground-based CMB experiments, including SPTPol\cite{Austermann2012} and SPT\cite{Sobrin2022}.
Mission success is driven by good overall availability of the instrument, and does not require perfect operation of every bolometer at all times.

The overall reliability of the readout system is determined by four factors:
\begin{itemize}
	\item The individual reliabilities of the subsystems comprising it,
	\item The manner in which they are interconnected (e.g. cross-strapping),
	\item The method in which the instrument is operated, and
	\item The quality required for data products to be considered ``good''.
\end{itemize}

Here, we focus only on the first factor, which is determined by the hardware itself and not mission-level architecture.
(In the case of LiteBIRD, SPUs are redundant, SCAs are not, and the ability to operate with failed SQUIDs or bolometers -- due to yield, aging, failure to tune, or any other reason -- is expected.)
A payload-level Fault Detection, Isolation, and Recovery analysis will provide payload-level reliability analysis and will be completed in an upcoming mission phase.

With 2 exceptions (see \Cref{sec:da}), components for all assemblies in the system are chosen from vendors' radiation-tolerant or radiation-hardened product lines.
The minimum total dose (TID) rating for all rad-tolerant/rad-hard components in the system is 20 krad.
Critical components within the system have higher total-dose ratings:
for example, the FPGA, switching, and linear power regulators are rated to a TID of 100 krad or higher.

Estimated failure rates for various architectural elements of the readout system are given in \Cref{tab:fit-rates}.
These failures were calculated by considering specific (where possible) or typical FIT rates for individual components.
Rates are exclusive of firmware errors, including radiation effects, which can interrupt operation but cannot cause permanent damage.
For the DA and SCA, we have separated FIT rates into failures affecting common circuitry used by the entire assembly (i.e. shared power supplies and control interfaces) and failures that can be isolated to only one channel (affecting only bolometers associated with a single SQUID).
These failures are treated separately because of their relative impacts.

\begin{table}
	\centering
	\begin{tabular}{l|l|r|r}
		\hline
		\textbf{Subsystem} & \textbf{Failure Type} & \textbf{FIT Rate} & \textbf{MTTF (Years)} \\ \hline
		\multirow{2}{*}{SCA} & single channel & 123 & 930 \\
		& common & 119 & 960 \\ \hline
		\multirow{2}{*}{DA} & single channel & 167 & 680 \\
		&common & 166 & 690 \\ \hline
		SPA & common & 510 & 220 \\ \hline
		PCA & common & 2,304 & 50 \\ \hline \hline
		\textbf{Complete SPU (SPA, PCA, 4 DAs)} & any & 6,150 & 18 \\ \hline
	\end{tabular}
	\vspace{2ex}
	\caption{
		Expected Failure-in-Time (FIT) Rates for individual architectural elements.
		FIT rates indicate expected failures per $10^9$ hours of operation.
		These values are calculated by considering specific (where possible) or typical FIT rates for individual components.
		FIT rates for mechanical/structural elements and cable assemblies are not included.
		The PCA FIT rate is known to be high relative to commercially available, space-qualified units and will be revised.
		The combined SPU FIT rate does not consider continued operation with failed units.
	}
	\label{tab:fit-rates}
\end{table}

\section{Next Steps}
\label{sec:next}

In this work, we described the major design elements and demonstration for a TRL-5 space-qualified readout system for TES detectors.
The next phase of development includes integrated full-scale functional testing, and a detailed characterization of noise performance with space-representative hardware.

\acknowledgments

The authors gratefully acknowledge the support of the Canadian Space Agency (CSA), through Phase-0 (9F050-190058/001/MTB) and STDP (9F063-190285/003/MTB) contracts.

\bibliography{hardware}
\bibliographystyle{spiebib}

\end{document}